\documentclass[aps,prl,twocolumn,superscriptaddress]{revtex4}
\usepackage{amsmath,amssymb}
\usepackage{graphicx}
\usepackage{bm}
\begin{document}

\preprint{LA-UR-08-05224}

\title{First order magnetic transition in single crystal CaFe$_2$As$_2$
detected by $^{75}$As NMR}


\author{S.-H. Baek}
\email[]{sbaek@lanl.gov}
\affiliation{Los Alamos National Laboratory, Los Alamos, NM 87545, USA}
\author{N. J. Curro}
\affiliation{Department of Physics, University of California, Davis, CA 95616, USA}
\author{T. Klimczuk}
\affiliation{Los Alamos National Laboratory, Los Alamos, NM 87545, USA}
\affiliation{Faculty of Applied Physics and Mathematics, Gdansk University of
Technology, Narutowicza 11/12, 80-952 Gdansk, Poland}
\author{E. D. Bauer}
\affiliation{Los Alamos National Laboratory, Los Alamos, NM 87545, USA}
\author{F. Ronning}
\affiliation{Los Alamos National Laboratory, Los Alamos, NM 87545, USA}
\author{J. D. Thompson}
\affiliation{Los Alamos National Laboratory, Los Alamos, NM 87545, USA}

\date{\today}

\begin{abstract}
We report $^{75}$As Nuclear Magnetic Resonance data in a single
crystal of CaFe$_2$As$_2$. The Knight shift, electric field gradient, and
spin-lattice relaxation rate are
strongly temperature dependent in the paramagnetic state, and change
discontinuously at the structural transition temperature, $T_S=T_N=167$ K.
Immediately below, the NMR spectra reveal an internal field at the As site
associated with the presence of a commensurate magnetic order. These results
indicate that the structural and magnetic transitions in CaFe$_2$As$_2$ are
first order and strongly coupled, and that the electron density in the FeAs
plane is highly sensitive to the out-of-plane structure.

\end{abstract}



\maketitle

The discovery of superconductivity in
LaFeAsO$_{1-x}$F$_{x}$ with $T_c=26$ K \cite{kamihara08} has attracted
interest due to structural and magnetic similarities with high-$T_c$ cuprates.
To date, much effort has been devoted to the search
for new iron-based compounds exhibiting an even higher $T_c$.  By replacing La
with other rare earths, such as Sm \cite{xchen08,ding08,ren08:2}, Ce
\cite{gchen08}, and Nd \cite{ren08}, $T_c$ has been raised to 55 K for Sm and
to 54 K in the oxygen deficient
$R$FeAsO$_{1-\delta}$ systems ($R$= Nd \cite{kito08}, Gd \cite{yang08}). In
both cases, the magnetic and structural transitions in the undoped parent material
are suppressed before entering the superconducting phase.
Further studies have shown that
the ternary FeAs compounds $A$Fe$_2$As$_2$ ($A$=Ba, Sr, Eu, and Ca) share
similar magnetic and structural properties as the $R$FeAsO parent compound \cite{rotter08:1,
ni08, jeevan08, zren08, ronning08}, and exhibit superconductivity by
doping $A$ with K or Na \cite{rotter08:2, ni08, wu08, sasmal08, gchen08:2}
or by applying pressure \cite{park08, tori08,alireza08} to suppress
the magnetic and the structural anomalies. 
These similarities suggest that the physics of both families of materials is 
dominated by FeAs layers and that `intercalated' layers serve primarily as 
tunable charge reservoirs.  

Because single crystals
of the ternary compounds grow more easily and have a simpler structure than
the quaternary compounds,
they appear to be an ideal
system to investigate the Fe-based superconductors.
They form in the well-known ThCr$_2$Si$_2$-type crystal
structure and undergo a spin-density wave transition
which accompanies a structural transition from tetragonal $I4/mmm$ to
orthorhombic $Fmmm$.  Neutron-diffraction studies find an ordered Fe moment of about
1 $\mu_B$ that develops along the orthorhombic $a$ axis with
antiferromagnetic (AFM) wave vector (1,0,1)
\cite{zhao08, jesche08, su08, kaneko08, goldman08}. 
Superficially, we might expect, then, that the relationship among structure, 
static magnetic order, and spin dynamics would depend only weakly on the 
isovalent $A$ atom. Establishing this expectation would provide a common 
framework for theoretical models of the parent compounds. As we will show, 
though, there are significant differences among the $A$Fe$_2$As$_2$ materials.

In this Letter, we present  $^{75}$As nuclear magnetic resonance (NMR) data in 
a single crystal of CaFe$_2$As$_2$.  In addition to providing unambiguous 
evidence for a first order spin-density wave (SDW) instability that 
occurs simultaneously with a first order structural transition, these studies 
show that, in contrast to most bulk measurements, the low
energy static and dynamic NMR properties (Knight shift, electric
field gradient (EFG), and $T_1^{-1}$) differ significantly from the
isostructural BaFe$_2$As$_2$ material.

\begin{figure}
\label{fig:1}
\centering
\includegraphics[width=0.48\textwidth]{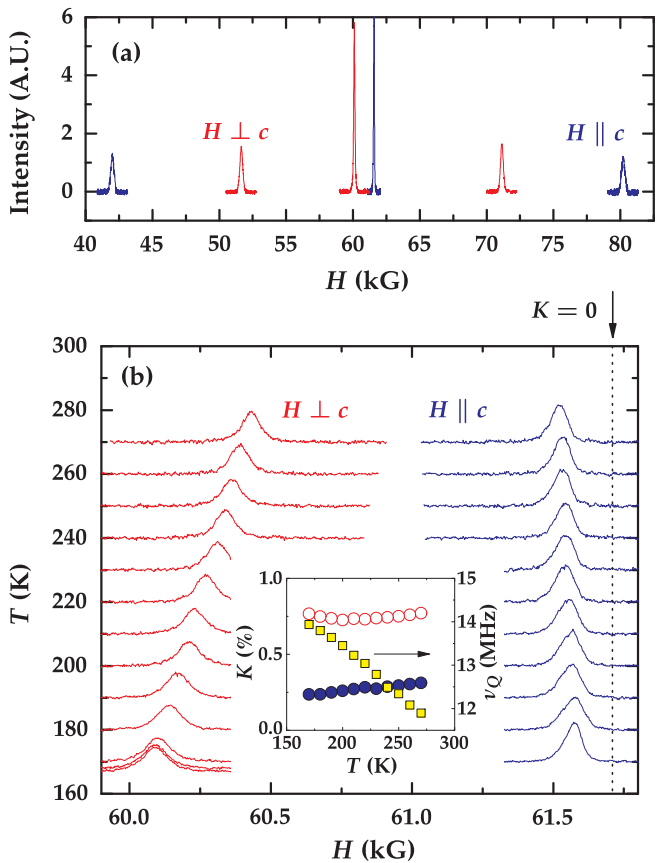}
\caption{$^{75}$As NMR spectra in the paramagnetic state at a fixed
frequency of 45 MHz.
(a) Full spectra with
satellites associated with both $H\perp c$ (red lines) and $H \parallel c$ (blue
lines) obtained at 170 K. (b) Central transition spectra for both field
orientations as a function of temperature.  For $H \perp c$ (red), the strong 
temperature dependence of $\nu_Q$ dominates the line position. The inset shows
$K$ vs.~$T$ for  $H \parallel c$ (blue) and $H \perp c$ (red), as well as 
$\nu_Q=\nu_c$ (yellow).} 
\end{figure}

Single crystals of CaFe$_2$As$_2$ (Ca122) were grown in Sn flux, using a
slightly different recipe than described in Ref.~\cite{ronning08}.
The starting elements
were placed in an alumina crucible and sealed under vacuum
in a quartz ampoule. The ampoule was placed in a
furnace and heated to 600 $^\circ$C at 100 $^\circ$C/hr, and held
at that temperature for 4 hours. This sequence was repeated
at 900 $^\circ$C and at a maximum temperature of 1075
$^\circ$C, with hold times of 4 hr, each. The sample was then
cooled slowly (7 $^\circ$C/hr) to 650 $^\circ$C, at which point the
excess Sn flux was removed with the aid of a centrifuge.
The resulting crystals, which form in the tetragonal ThCr$_2$Si$_2$ structure 
that can be viewed as layers of Ca capped by Fe-As tetrahedra along the 
$c$-axis, exhibit a first order transition at 167 K, which is slightly lower than 171 K
in Ref.~\cite{ronning08}.
This may indicate that the transition temperature is affected weakly by
subtle changes in the growth condition or by the
exact amount of substitutional Sn that is incorporated into the crystal from the Sn
flux out of which crystals grow.
However, Ca122 seems to tolerate little Sn doping,
unlike BaFe$_2$As$_2$ in which the transition temperature is suppressed to 85
K from 140 K in Sn-free samples \cite{ni08}.
Regardless of the slightly lower transition temperature, the magnetic
susceptibility $\chi(T)$ shows similar temperature and field dependences, and
the resistivity data confirm the same anomaly at 167 K and its thermal
hysteresis as previously reported \cite{ronning08},
indicating comparable quality of these single crystals.

Fig.~1(a) shows NMR spectra of $^{75}$As ($I=3/2$) at 170 K and at a fixed
resonance frequency of 45 MHz,
for both $H \parallel c$ (blue lines) and $H \perp c$ (red lines). The spectra
are fit well by a nuclear Hamiltonian:
$\mathcal{H}=\gamma\hbar(1+K_{\alpha})\hat{I}_{\alpha}H_0 + h\nu_c/6
[(3\hat{I}_c^2-1) + \eta(\hat{I}_a^2 - \hat{I}_b^2))]$,
where $K_{\alpha}$ is the magnetic shift in the $\alpha$ direction,  ${a,b,c}$
are the unit cell axes,
$\nu_c$ is the EFG in the $c$ direction, $\eta$ is the anisotropy factor, and
the NQR frequency is given by $\nu_Q = \nu_c\sqrt{1+\eta^2/3}$. We find that
$\nu_c=13.93$ MHz and $\eta=0$ with the principal axis of the EFG tensor along
the $c$-direction in the paramagnetic state at $T=170$ K. This value is nearly
500 \% larger
than $\nu_Q$ measured in BaFe$_2$As$_2$ \cite{baek08,kitagawa08}. By measuring
the temperature dependence of the satellite transition
($I=+\frac{3}{2}\leftrightarrow+\frac{1}{2}$) for $H \parallel c$ (not shown), we extract 
the temperature dependence of $\nu_c(T)$, shown in the inset of Fig.~1(b).  The
EFG increases by 16\% between room temperature and $T_N$. This behavior
contrasts sharply with that observed in BaFe$_2$As$_2$, where $\nu_c(T)$
decreases by the same amount over the same temperature range as shown in the
inset of Fig.~3.   The EFG at the
As site is given by the sum of a lattice term 
($\nu_c^\text{lattice} \propto 1/V_\text{cell}$) and an on-site term
$\nu_c^\text{on-site}$. The changes in $\nu_c$ we observe far exceed the change
of the unit cell volume $V_\text{cell}$ between both compounds and the lattice
contraction over this range of temperature \cite{ronning08,ni08:2}, therefore the dominant
contribution to the EFG must be on-site charge distribution in the As $4p$
orbitals. In contrast with the cuprates, our results indicate that
the charge distribution in the FeAs planes changes dramatically from one
material to the other, and probably reflects the sensitivity of the ground
state to pressure. In fact, pressure-induced
superconductivity is found at the relatively modest
pressure of 0.4--0.8 GPa in CaFe$_2$As$_2$ compared to 2.8--3.5 GPa in SrFe$_2$As$_2$ and
2.5--5.5 GPa in BaFe$_2$As$_2$ \cite{park08,tori08,alireza08}. These results may 
reflect different amounts of charge donation from the ionic layer. 

The temperature dependences of the central transition in the paramagnetic (PM)
state are shown in Fig.~1 (b). The Knight shift ($K$) reveals a strong anisotropy of the
spin susceptibility, as shown in the inset of Fig.~1 (b).
Like  BaFe$_2$As$_2$,  $K_{ab}>K_c$ suggests that the spin susceptibility is
greater in the plane, which is also the case for
$\mathrm{LaO_{0.9}F_{0.1}FeAs}$ \cite{grafe08}.  In
contrast, however, we find that $K_{ab}$ exhibits a shallow upturn just above
$T_N$.  The origin of this behavior is not understood. We have not attempted 
to extract the hyperfine coupling in CaFe$_2$As$_2$ since the susceptibility 
shows a strong paramagnetic impurity contribution. 

\begin{figure}
\label{fig:2}
\centering
\includegraphics[width=0.48\textwidth]{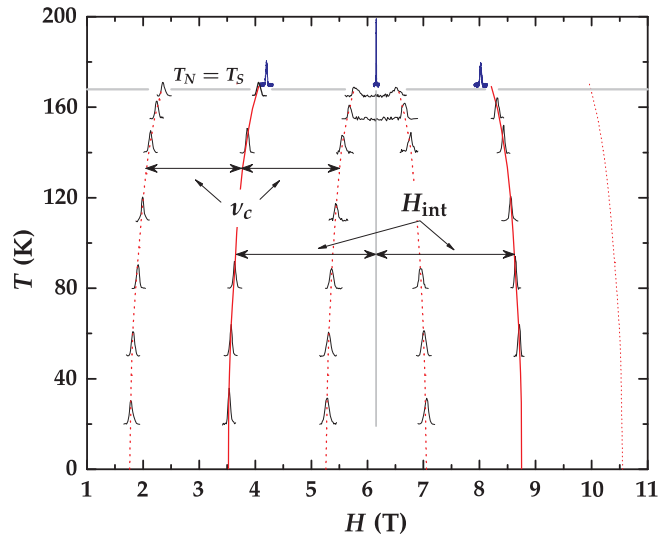}
\caption{Temperature dependences of $^{75}$As NMR spectra below the transition
for $H \parallel c$.  The spectra (blue lines) in the paramagnetic state split 
to six lines by the internal field $H_\text{int}$ in the ordered state. The red
solid and dotted lines represent the splited central lines and the satellites 
associated with each central line, respectively. One of satellites at highest fields was 
not measured due to the limited maximum field (9 T) in our magnet.  
Horizontal line (gray) denotes $T_N=T_S=167$ K. } 
\end{figure}

At 167 K, we observe an abrupt change of the spectrum, as shown in Fig.~2.
Both the central and satellite resonances are split by an internal field
$H_\text{int}$ as a result of the hyperfine coupling between the As nuclei and
the ordered Fe moments.  Since the central line is split into two resonances
rather than simply shifted to lower field, we conclude that
$\mathbf{H}_\text{int}$ is either parallel or antiparallel to $\mathbf{H}$, the
applied field.  In this case, the resonance fields are given by
$H_\text{central} = \nu_0/\gamma \pm H_\text{int}$ and
$H_\text{sat} = (\nu_0-\nu_c)/\gamma \pm H_\text{int}$. The temperature dependences of
$\nu_c(T)$ and $H_\text{int}(T)$ are shown in Fig.~3.  We find that
$H_\text{int} = 2.6\pm 0.1$ T, which is a factor of two larger than the value of 1.3 T 
observed in BaFe$_2$As$_2$ \cite{kitagawa08}. Furthermore, we see only one value of
$|H_\text{int}|$, indicating a commensurate magnetic structure.  If the magnetic
structure were incommensurate with the lattice, then the internal field would
be distributed and the spectrum would not exhibit the sharp resonances seen in
Fig.~2.  Recent neutron scattering results are consistent with our data
\cite{goldman08}.

We also observe a discontinuous decrease in $\nu_c(T)$ at $T_N$, which is very 
similar to the case in BaFe$_2$As$_2$ (inset, Fig.~3), although the value of 
$\nu_c$ and its temperature dependence in the PM state is clearly different. 
The reason for the difference in $\nu_c$ between these two isostructural compounds is
unclear, but may reflect the extreme sensitivity of the electronic
structure to the out-of-plane atoms.  
Clearly, both the magnetic order
parameter, given by $H_\text{int}(T)$, and a measure of the structural distortion,
given by $\Delta\nu_c(T) = |\nu_c(T) - \nu_c(T_N)|$, are discontinuous at $T_N$,
indicating the first-order nature of the transition in CaFe$_2$As$_2$.
Upon warming the sample from the ordered state, the paramagnetic signal is
recovered at 168 K, revealing a thermal hysteresis of 1 K in excellent
agreement with
results from neutron diffraction \cite{goldman08}.
We emphasize that
there is no temperature range in which we observe either the magnetic or
structural order parameter finite and the other one zero, indicating that both
are intimately related.  The temperature dependence of 
$H_\text{int}$ observed in
Fig.~3 is remarkably close to the temperature dependence of the ordered moment 
that develops below a first order magnetic transition in isostructural SrFe$_2$As$_2$ 
\cite{kaneko08}.

\begin{figure}
\label{fig:3}
\centering
\includegraphics[width=0.48\textwidth]{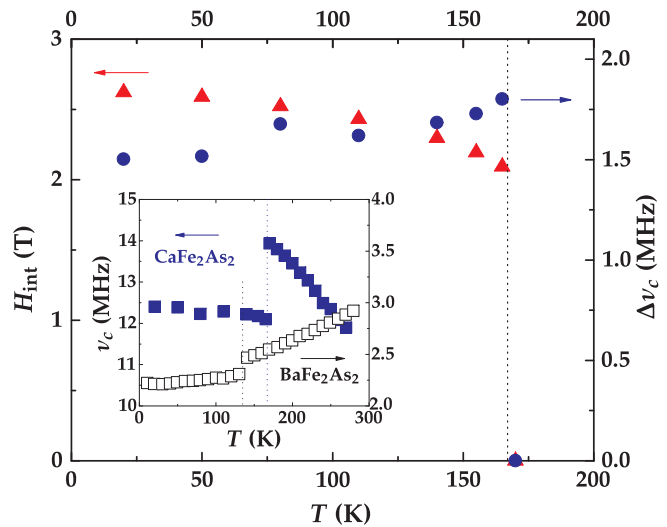}
\caption{Temperature dependences of the order parameters obtained from NMR
spectra in the ordered state for $H \parallel c$.  $H_\text{int}$ is 
proportional to the sublattice magnetization, and is a measure of the magnetic 
order parameter, while $\Delta\nu_c \equiv |\nu_c(T)-\nu_c(T_N)|$ is 
a measure of the structural distortion.  The inset shows the 
temperature dependences 
of $\nu_c (T)$ for both CaFe$_2$As$_2$ and BaFe$_2$As$_2$ (the latter is 
reproduced from Ref.~\cite{kitagawa08}).}
\end{figure}

The relationship between $H_\text{int}$, the ordered moments $\mathbf{S}_0$,
and the magnetic structure is not straightforward.  \textit{A priori}, one
might expect the hyperfine field to vanish at the As site due its symmetric
position between the four nearest neighbor Fe sites.  However, this is the
case only if the transferred hyperfine coupling to the As atom is isotropic.  Kitagawa
et al.~\cite{kitagawa08} report a model for the hyperfine coupling in terms of anisotropic
coupling tensors $\mathbf{B}$ between the four nearest neighbor Fe moments and
the As nucleus.  In this case, $H_\text{int} = 4B_{ac} S_0^x$ for the
$\mathbf{Q}=(101)$ stripe magnetic structure \cite{goldman08}.  Since the
neutron scattering data reveal $S_0 = 0.8 \mu_B$ oriented along the 100
direction, we estimate $B_{ac} \sim 0.81$ T/$\mu_B$.  In this case, the
transferred hyperfine coupling must be anisotropic in order to induce a
hyperfine field.  A second possibility is that the ordered moments are canted
by the applied field and acquire a small component along the $c$ direction, 
for which the isotropic component of the transferred hyperfine coupling
does not vanish.  If we use the isotropic values reported for BaFe$_2$As$_2$ 
(2.64 T/$\mu_B$) \cite{kitagawa08}
and an ordered moment of 0.8 $\mu_B$, then we find that the moments must be
tilted by $\sim 22.5^{\circ}$ out from the $ab$ plane.

\begin{figure}
\label{fig:4}
\centering
\includegraphics[width=0.48\textwidth]{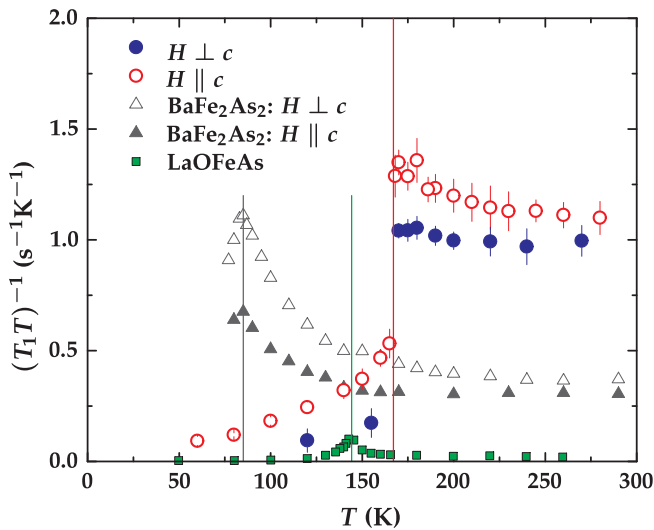}
\caption{$(T_1T)^{-1}$ for CaFe$_2$As$_2$, BaFe$_2$As$_2$ (Ref.~\cite{baek08}) and LaOFeAs 
(reproduced from \cite{nakai08}) as a function of $T$. The data reveal a 
discontinuity at $T_N$ and the formation of a gap at the Fermi level 
due to the SDW instability, and supports the first-order character of the 
magnetic transition. Clearly, the spin dynamics in the paramagnetic state is a 
strong function of the particular material.} 
\end{figure}

The nuclear spin-lattice relaxation rate ($T_1^{-1}$) was determined by
fitting the recovery of the nuclear magnetization using a Hahn-echo sequence
after a saturating pulse.
$(T_1T)^{-1}$ is shown as in Fig.~4.  At high temperatures $T\gg T_N$,
$(T_1T)^{-1}$ approaches a constant value, as observed in BaFe$_2$As$_2$ \cite{baek08} and
$R\mathrm{O}_{1-x}\mathrm{F}_x\mathrm{FeAs}$ ($R$=La,Pr) \cite{nakai08,matano08}.  
This Korringa-like
behavior is expected in metallic systems, and may reflect the coupling of the
nuclei to the conduction electrons.  With decreasing temperature, $(T_1T)^{-1}$
increases as $T_N$ is approached.  We attribute the upturn in
$(T_1T)^{-1}$ to dispersive (paramagnon) excitations that recent neutron 
scattering experiments find at temperatures well above $T_N$ ($\sim200$ K) \cite{mcqueeney08}. 
At $T_N$ we observe a discontinuous jump in $(T_1T)^{-1}$ in both field directions,
providing further evidence
for the first-order character of the magnetic transition. Below $T_N$, $T_1^{-1}$ decreases
exponentially with decreasing temperature. Upon further cooling, $(T_1T)^{-1}$
approaches a constant value, suggesting a partially gapped density of states at the
Fermi level that is expected for a SDW ground state.  Qualitatively, all
three compounds, CaFe$_2$As$_2$, BaFe$_2$As$_2$ and
$\mathrm{LaOFeAs}$ exhibit similar spin lattice
relaxation behavior, yet the absolute values of $(T_1T)^{-1}$ differ
dramatically.  This difference is surprising since the As probes the spin
fluctuations in similar FeAs planes in all three cases. There are two possible
explanations for this difference: either (i) the hyperfine coupling between
the Fe and the As changes between compounds, or (ii) the spectral density of
spin fluctuations changes.  However, the hyperfine coupling extracted from
plots of $K$ versus $\chi$ are roughly identical in
$\mathrm{LaO}_{1-x}\mathrm{F}_x \mathrm{FeAs}$ and BaFe$_2$As$_2$ \cite{kitagawa08,grafe08}.
Therefore, we conclude that the spectral density of spin fluctuations differs
significantly between these compounds.  One might argue that the single plane
${\rm La} {\rm O}_{1-x} {\rm F}_{x} {\rm Fe} {\rm As}$ should exhibit
different physics than the double plane $A$Fe$_2$As$_2$ compounds, but
apparently the spin fluctuations even differ for different $A$ atoms.  This
result points to the extreme sensitivity of the low energy excitations in
these materials to the particular structure of the out of plane atoms and the
external pressure.

In conclusion, we have found that the magnetic and structural transitions
occur simultaneously at 167 K in single crystal CaFe$_2$As$_2$. The
antiferromagnetic transition is clearly first order and commensurate. Also, 
the discontinuous formation of the gap associated with a spin-density wave 
instability at 167 K was directly demonstrated by $T_1^{-1}$ 
measurements. Comparison with isostructural BaFe$_2$As$_2$
and another parent compound LaOFeAs, 
demonstrates the extreme sensitivity of both the static ($\nu_Q$) and the 
dynamic ($(T_1T)^{-1}$) properties to the out-of-plane structure. 
Understanding this sensitivity and its ultimate connection to 
superconductivity may shed light on the optimal microscopic conditions 
for the highest $T_c$.

We thank Stuart Brown, Tuson Park, and Hanoh Lee for useful and delightful discussions.
Work at Los Alamos National Laboratory was performed under the
auspices of the US Department of Energy, Office of Science.

\bibliography{CaFe2As2_NMR}

\end{document}